\documentclass[10pt,journal,compsoc]{IEEEtran}
\usepackage{graphicx}
\usepackage{amsmath}
\usepackage{algorithm}
\usepackage{algpseudocode}

\usepackage{cite}
\usepackage{etoolbox}
\usepackage{float}
\usepackage{array, tabularx, caption}
\usepackage{cellspace}
\usepackage[font={small}]{caption}
\usepackage{subcaption}
\usepackage{cmap}
\usepackage[T1]{fontenc}

\usepackage[math]{blindtext}
\usepackage{cite}
\patchcmd{\thebibliography}{{\bibname}}{}{}{} 
\usepackage[utf8]{inputenc}
\usepackage[english]{babel}
\usepackage{xcolor}
\usepackage{colortbl}
\usepackage{booktabs}
\usepackage{pdfcomment}
\usepackage{xspace}

\begin{document}

\title{Proactive DDoS Attack Mitigation in Cloud-Fog Environment using  Moving Target Defense}

\author{\IEEEauthorblockN{Vaishali Kansal}\\
\IEEEauthorblockA{Department of Computer Science and Engineering\\
Indian Institute of Technology, Roorkee,
Uttarakhand, India\\
veshukansal@gmail.com}\\
\and
\IEEEauthorblockN{Mayank Dave}\\
\IEEEauthorblockA{Department of Computer Engineering\\
National Institute of Technology, Kurukshetra,
Haryana, India\\
mdave@nitkkr.ac.in }}

\maketitle

\begin{abstract}

Distributed Denial of Service (DDoS) attacks are serious cyber attacks and mitigating DDoS attacks in cloud is a topic of ongoing research interest which remains a major security challenge. Fog computing is an extension of cloud computing which has been used to secure cloud. Moving Target Defense (MTD) is a newly recognized, proactive security defense that can be used to mitigate DDoS attacks on cloud. MTD intends to make a system dynamic in nature and uncertain by changing attack surface continuously to confuse attackers. In this paper, a novel DDoS mitigation framework is presented to support  Cloud-Fog Platform using MTD technique (\textit{CFPM}). \textit{CFPM} applies migration MTD technique at fog layer to mitigate DDoS attacks in cloud. It detects attacker among all the legitimate clients proactively at the fog layer and isolate it from innocent clients. \textit{CFPM} uses an effective request handling procedure for load balancing and attacker isolation procedure which aims to minimize disruption to cloud server as well as serving fog servers. In addition, effectiveness of \textit{CFPM} is evaluated by analyzing the behaviour of the system before and after attack, considering different possible scenarios. This approach is effective as it uses the advantage of both MTD technique and Fog computing paradigm supporting cloud environment.

\end{abstract}


\section{Introduction}
\par Amongst the diverse threats to cyber security Distributed denial-of-service (DDoS) attack \cite{Koten,bb,Zar} is most prevailing nowadays. In DDoS attacks, target is overloaded with enormous traffic by a large number of connected devices that are distributed across the Internet. This results in loss of equipment resources and network bandwidth.
\par Securing cloud from these attacks has become a big challenge \cite{Beh,Yus,Somani}. Cloud servers have the potential to perform computing and storage tasks, without the direct involvement of user. But as they are faraway, it is required to send all the raw data (could be sensitive) to the cloud servers over the internet which may have some privacy and security issues. Various DDoS attack handling techniques in cloud \cite{Quan, WangCloud, Ahmed} are proposed.
\par Fog computing or fogging \cite{Vaq}, is a decentralized computing infrastructure that distributes data, services of computing, storage, networking and communications between the cloud and the data source. Fog computing is basically an extension of cloud computing. In order to provide faster ICT services to the end users, it brings cloud computing capabilities to the edge of the network. The goal of fogging is to enhance efficiency by reducing the total amount of sensitive data that is transported to the cloud for analysis, storage and processing \cite{Stol}.
\par Though fog computing conserves network bandwidth and improves system response time but at the same time it is also prone to cyber attacks such as DDoS attacks. As the capacity of each fog node is restricted, these attacks can be detrimental to fog computing's availability also \cite{Stog}.  As of now it is critical to secure cloud with additional measures as fog layer alone is not sufficient enough.
\par One newly recognized cyber defense strategy, which helps in mitigating DDoS attacks is introduced i.e. Moving Target Defense (MTD) \cite{Mtd, Zhuang,Li}. MTD keeps the network environment less deterministic and unpredictable by continuously changing the attack surface.  This uncertainity induces confusion for attackers, simultaneously increasing the attacker’s time, efforts, and cost while decreasing the cost of defensive strategies. 
MTD has its potential applications in various areas.
In the literature Fog computing and MTD has been used separately for cloud security. Also MTD techniques have not been applied to fog computing paradigm.
\par In this paper an approach \textit{CFPM} is presented, which mitigates DDoS attacks in Cloud-Fog Platform using Moving Target Defense(MTD) as the underlying technique. \textit{CFPM} monitors the traffic generated from all the connected users to look for abnormalities. It handles the DDoS attack at fog layer using the concept of migration MTD. Migration MTD is a dynamic platform diversification MTD technique. This approach is effective as it takes the advantage of both Fog computing and MTD to secure cloud.

\section{Related Work}
\par Several techniques have been developed to mitigate the impact of DDoS attacks in cloud \cite{ Quan, WangCloud}. Ahmed et al. \cite{Ahmed} investigated such mitigation techniques and compared them in cloud environment. Fog computing \cite{Vaq} is one of the promising way to deal with such attacks in cloud. Decoy information technology called Fog computing has been used by the authors in \cite{Stol} to secure the cloud. But fog computing itself is susceptible to such attacks, which is investigated by the authors in \cite{Stog} as fog devices work at the bottom/edge of networks.
\par MTD emerged as a promising approach to network security. Li Jason et al. \cite{Li} productized MTD research technology, called Self-shielding Dynamic Network Architecture (SDNA) technology, now known as Cryptonite NXT. MTD intends to create asymmetric uncertainty on the attacker’s side by constantly changing the attack surface. Zheng et al.\cite{Zhenga} presented a comprehensive survey on MTD and its implementation strategies from architectural perspective. 
\par Based on the kind of changes MTD is making to the system it has been classified into four categories i.e., Software based diversification , Runtime based diversification, Communication diversification and Dynamic Platform diversification by the authors in \cite{V3}. MTD has its potential applications in various areas. Different MTD techniques have been applied in different domains. Authors in \cite{Kc,Sha} have applied runtime based diversification MTD techniques i.e., instruction-set randomization and address-space randomization to counter attack. Jackson et al. \cite{Jac} have used software based diversification MTD techniques i.e., software diversity and compiler generated diversity and evaluated their effectiveness. Al- Saher et al. \cite{Ai} presented a MTD architecture called mutable networks(MUTE) which uses communication diversification MTD technique i.e., IP address randomization. Thompson et al. \cite{Thom} proposes an approach that applies  rotation of multiple operating systems to enhance security i.e. diversity technique has been used which comes under dynamic platform diversification MTD.
\par Zhuang et al. \cite{Rui} applied simulation-based experiments to explain enhanced efficiency due to the introduction of diversification and randomization in the system. Comparison of different MTD techniques has been proposed by the authors in \cite{Xu}. Authors in \cite{Jin, Marco} studied the effectiveness of MTD using security models and investigated applications of MTD to network security.   
Okhravi et al. \cite{Okh,Ravi} has applied dynamic platform diversification MTD technique as defense mechansim on critical infrastructure. MTD has been used for mitigation of DDoS attacks against a proxy based architecture by the authors in \cite{Jia, Wang, Wood, Ven}. Authors in \cite{V1, V2} have used dynamic platform diversification i.e. shuffle MTD as the underlying MTD technique to mitigate DDoS attack proactively at proxy layer.
\par Peng et al. \cite{Peng} ascertained whether MTD is effective in protecting Cloud-based services and if yes, upto what extent, with heterogeneous and dynamic attack surfaces. Wang et al. \cite{W} proposed a cost effective defense against DDoS and Covert channel attacks using MTD. Authors in \cite{HM} have used IP address randomization as MTD technique for cloud computing platform. The authors in \cite{Hoo,Alav} have combined differnt dynamic platform diversification MTD techniques i.e, Shuffle and Diversity and evaluated it for cloud computing. This combined approach is very efficient as it amplifies randomization. So due to its promising capabilities MTD is a suitable strategy for addressing cloud computing security issues. But MTD has not been applied to fog paradigm in the field of research in past. Bazm et al.\cite{Moh} evaluated that MTD can be applied to both multi-cloud and decentralized cloud infrastructures such as Fog. 
\par This paper presents an approach \textit{CFPM} that mitigates DDoS attack in a combined Cloud-Fog Platform using dynamic platform diversification MTD technique i.e., migration MTD technique.

\section{System Model}
\subsection{Overview}
\par
\begin{figure}[h]
\centering
\includegraphics[width=3in,height=3.2in]{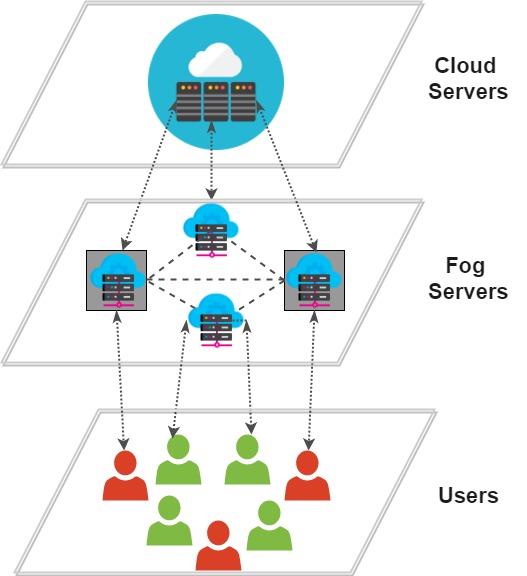} 
\caption{System Model}
\label{v1}
\end{figure}

 The proposed approach \textit{CFPM}, mitigates DDoS attacks in the cloud by using the concept of fog computing with migration moving target defense technique. Fig. \ref{v1} shows the proposed architecture, which consists of 3 components: 
\begin{enumerate}
\item(a) cloud servers
\item(b) fog layer comprising of many fog servers
\item(c) users (both innocent and malicious)
\end{enumerate}

\par Fog servers present in the fog layer are in two modes i.e. sleep mode or active mode. Active fog servers are used for serving the users and the grey colored fog server i.e., sleeping fog server is activated only if any attacker is detected so sleeping fog server is named as Attack Fog server. Number of sleeping fog servers are very less compared to active fog servers. 
\par This architecture permits the user to access the services provided by fog servers only after successful authentication. User before connecting to any fog server mention the maximum requests it can send per second and at the time of requesting it passes its current need. 
\par A fog server can serve many users and a user can request any number of times in a second until or unless it has reached its limit i.e., it has requested the maximum number of requests in total that it has mentioned at the time of connection. If the request can not be handled at the fog layer i.e., all fog servers are fully occupied, then only the request is passed to the cloud server.

\subsection{Preliminaries}

\par  Table \ref{tab:tablec} shows various notations used in the \textit{CFPM} approach presented in this paper.

\begin{table}[h]
\centering
\caption{Main Notations}
\label{tab:tablec}\centering
\begin{tabularx}{8cm}{|Sl|X|}
\hline
\textbf{Notation} & \textbf{Meaning} \\
\hline
\hline
uid & user id \\
\hline
fid & fog id \\
\hline
 thresh & data processing speed of fog server \\
 \hline
  $F_{cap}$ & maximum requests a fog server can handle per second\\
 \hline
 $N_{usr}$ & total number of users \\ 
 \hline
 $Req_{sz}$ & size of the request \\
 \hline
 max[i] & maximum requests that a user "i" can send per second \\
 \hline
 limit[i] & maximum usage of a user "i" \\
 \hline
countfid[i] & number of requests served by a fog server for a user "i" \\
\hline
flag[i] & state of each fog server "i" \\
\hline
account[i] & number of requests handled for a particular user "i" by either one fog server or a number of fog servers or cloud server collectively\\
\hline
score[i] & current usage of user "i" \\
\hline
\end{tabularx}
\end{table}

\par
\begin{figure}[h]
\centering
\includegraphics[width=3.8in,height=1.5in]{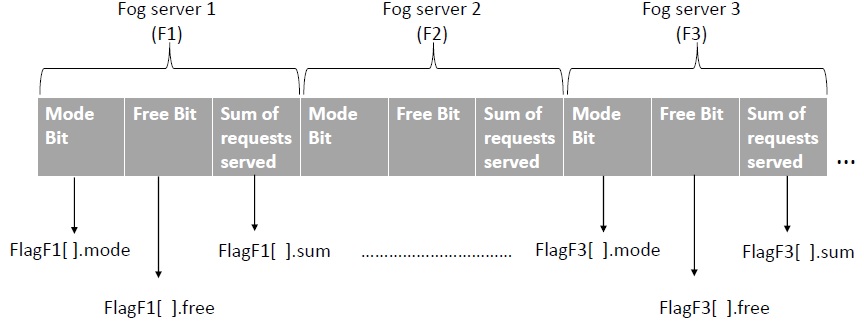} 
\centering
\caption{Structure of Flag[ ] variable}
\label{v2}
\end{figure}

 Fig. \ref{v2} shows the structure of \emph{Flag[ ]} variable, which defines the state of each fog server using 3 fields as follows:

\begin{enumerate}
\item Mode Bit: 
\begin{itemize}
\item If set to 0, then it signifies the fog server is in sleep mode.
\item If set to 1, then it signifies the fog server is in active mode.
\end{itemize}
\item Free Bit:
\begin{itemize}
\item If set to 0, then it signifies the fog server is fully occupied.
\item If set to 1, then it signifies the fog server can serve more requests.
\end{itemize}
\item Sum: It is the sum of the entries in \emph{countfid[ ]} array maintained for each fog server i.e., the total number of requests served by a fog server.
\end{enumerate}

\section{Proposed Scheme}
In this section details of the proposed scheme are discussed.
\subsection{Initial Assignment}
\par For the initial assignment of all the local and global variables used in \textit{CFPM} approach, Algorithm 1 is executed. 
 \begin{algorithm}[h]
\centering
\caption{Initial Assignment}
\begin{algorithmic}[1]
\Function{Global-Assign}{}
\For  { int i = 1 to $N_{usr}$ }
\State account[i] = 0;
\State max[i] = max requests user can send per second;
\State limit[i] = $max[i] \times Req_{sz}$
\EndFor
\For {every fid}
\State Flag[fid].sum = 0;
\State Flag[fid].free = 1;
\EndFor
\EndFunction
     
\Function{Local-Assign}{}
\For { every fid}
\For  { int i = 1 to $N_{usr}$ }
\State countfid[i] = 0;
\EndFor
\EndFor

\EndFunction
\end{algorithmic}
\end{algorithm}
\par After each second GLOBAL-ASSIGN() and LOCAL-ASSIGN() functions are called to initialize the variables. For all fog servers mode bit in \emph{Flag[ ]} variable is initialized at the time of fog server creation.

\subsection{Request Sending Procedure}
\par Whenever any user sends a request to a fog server or a fog server sends a request to another fog server, then Algorithm 2 is executed. 
\begin{algorithm}[h]
 \centering
   \caption{Send Request}
    \begin{algorithmic}[1]
   
      \Function{Send-Request}{uid, need}
      \State For any fid
      \If {Flag[fid].mode = = 1 \&\& Flag[fid].free = = 1}
      \If {user request}
      \State HANDLE-REQUEST(uid, need);
      \Else 
      \State SERVE(uid, need);
      \EndIf
      \Else 
      \State Forward Request-CloudServer(uid, need);
      \EndIf
 \EndFunction
\end{algorithmic}
\end{algorithm}

\par Firstly it is checked whether there is any active free fog server available or not by accessing the mode bit and free bit of \emph{Flag[ ]} variable for all fog servers. If for all fog servers free bit is 0 i.e., no fog server is free to serve the request, then the request is send to the cloud server.
\par Otherwise if for any fog server mode bit and free bit is 1, then it is checked whether the request is from user or neighboring fog server. If request is from user, then HANDLE-REQUEST(uid, need) function is called in which the user id of the user to be served is passed along with its current need. Otherwise if the request is from another fog server, then SERVE(uid, rem) function is called and, the user id of the user is passed for which the requested fog server was unable to serve along with the remaining need of the user.

\subsection{Request Handling Procedure}
 \par Whenever any fog server receives a request from a user or any other neighboring fog server it runs Algorithm 3.
 
  \begin{algorithm}[h]
 \centering
   \caption{Request Handling}
    \begin{algorithmic}[1]
      \Function{Handle-Request}{uid, need}
      \If {account[uid] != NULL}
      \State account[uid] = account[uid] + need;
      \State score[uid] = $account[uid] \times Req_{sz}$
      \If {score[uid] $>$ limit[uid] }
      \State ACTIVATE-ATTACKFOG(uid);
      \Else 
      \State SERVE(uid, need);
      \EndIf
      \EndIf
       \EndFunction
       
      \Function{Serve}{uid, need}
      \State countfid[uid] = countfid[uid] + need;
      \State int sum = 0;
      \For {int i = 1 to $N_{usr}$}
      \State sum = sum + countfid[i];
      \EndFor
      \If { sum = = $F_{cap}$ }
     
           \State  Flag[fid].sum = sum;
            \State Set Flag[fid].free = 0;
           
  \ElsIf { sum $>$ $F_{cap}$ }
  
  \State rem = sum - $F_{cap}$;
  \State countfid[uid] = countfid[uid] - rem; 
  \State Flag[fid].sum = $F_{cap}$;
  \State Set Flag[fid].free = 0;
  \State SEND-REQUEST(uid, rem);
  \State exit( );

            \Else
            \State Flag[fid].sum = sum;
            \EndIf
            \EndFunction
            
\end{algorithmic}
\end{algorithm}
\par If the request is from user, then HANDLE-REQUEST(uid, need) function is called otherwise if the request is from another fog server, then only SERVE(uid, need) function is called.
 \par In HANDLE-REQUEST(uid, need) function the \emph{account[uid]} value of the user to be served is accessed. If the value is not null i.e, the user is not present in blacklist, then the fog server updates the \emph{account[uid]} value by adding the current need of the user into it. Then it calculate the \emph{score[uid]} value using Eq. \ref{equ1}.
 \begin{equation}\label{equ1}
score[uid] = account[uid] \times Req_{size}
\end{equation} 
\par It then compares the \emph{score[uid]} value with the \emph{limit[uid]} value. If \emph{score[uid]} value is greater than the \emph{limit[uid]} value, then ACTIVATE-ATTACKFOG(uid) function is called by the fog server to deal with the attacker who is trying to send more requests than its maximum requests. Otherwise if \emph{score[uid]} value is less than or equal to the \emph{limit[uid]} value, then fog server accepts the user request and calls SERVE(uid, need) function.
\par In SERVE(uid, need) function fog server updates the \emph{count[uid]} value by adding the need value. It then calculates the total number of requests it is currently serving i.e., sum of the entries of its \emph{countfid[ ]} variable. 
\par All the fog servers have the same data processing speed so maximum requests it can serve per second i.e., $F_{cap}$ would also be same that can be calculated using Eq. \ref{equ2}.
\begin{equation}\label{equ2}
F_{cap} = \frac {thresh}{Req_{size}}
\end{equation}
 \par A fog server is performing many functions so to prevent it from getting overloaded, the number of requests that it can serve should be restricted to a certain number. We are assuming that all the fog servers have a certain capacity $F_{cap}$  depicting the number of requests it can serve in total. Using this constraint the \textit{CFPM }approach balances the load among all the available fog servers.
 \par Further, the fog server compares its capacity with the total requests it has served. If its capacity has reached to its \emph{limit} value i.e., if sum value is equal to $F_{cap}$, then it updates \emph{Flag[fid].sum} equal to sum and sets its \emph{Flag[fid].free} value to 0 and exit. 
 \par Otherwise if sum value is greater than $F_{cap}$, then it means the fog server is unable to serve the user completely. It then calculate the remaining number of requests and updates the \emph{count[uid]} value equal to the number of requests it has served for the user. It updates \emph{Flag[fid].sum} equal to $F_{cap}$ and sets its \emph{Flag[fid].free} value to 0. Further, it calls the SEND-REQUEST(uid, rem) function to request to another available fog server to serve the remaining need of the user.
 \par But if sum is less than $F_{cap}$, then it signifies the normal execution where the fog server is able to serve the user completely or serves its remaining need completely. So in this it only updates \emph{Flag[fid].sum} equal to sum.

\subsection{Attacker Isolation Procedure}
\par The proposed approach \textit{CFPM}, uses Attack Fog servers to handle attack. Algorithm 4 is executed if ACTIVATE-ATTACKFOG(uid) function is called by any fog server.
\begin{algorithm}[h]
 \centering
   \caption{Attack Fog Activation and Attacker Isolation}
    \begin{algorithmic}[1]
      \Function{Activate-AttackFog}{uid}
      \State Activate any fog server with Flag[fid].mode = 0;
      \State Flag[fid].mode = 1;
      \State ATTACKER-ISOLATION(uid, fid);
       \EndFunction
       \Function{Attacker-Isolation}{uid, fid}
         \State uid $\gets$ attacker;
         \State Add uid to blacklist;
         \State account[uid] = NULL;
        \State Assign uid to fid;
      \State Flag[fid].mode = 0;
       \EndFunction
\end{algorithmic}
\end{algorithm}

\par In this a sleeping fog server having \emph{Flag[fid].mode} as 0 is activated by changing \emph{Flag[fid].mode} value to 1. Further, ATTACKER-ISOLATION(uid, fid) function is called in which user id received in the function is declared as attacker and is added to the blacklist. The \emph{account} value of the attacker is set to NULL to signify that it would no longer be served by any fog server. 
\par After that the attacker is migrated to the attack fog server and the activated fog server again sets its \emph{Flag[fid].mode} value to 0, which results in isolation of the attacker.

\begin{figure}[H]
\centering
\includegraphics[width=3.8in,height=4.2in]{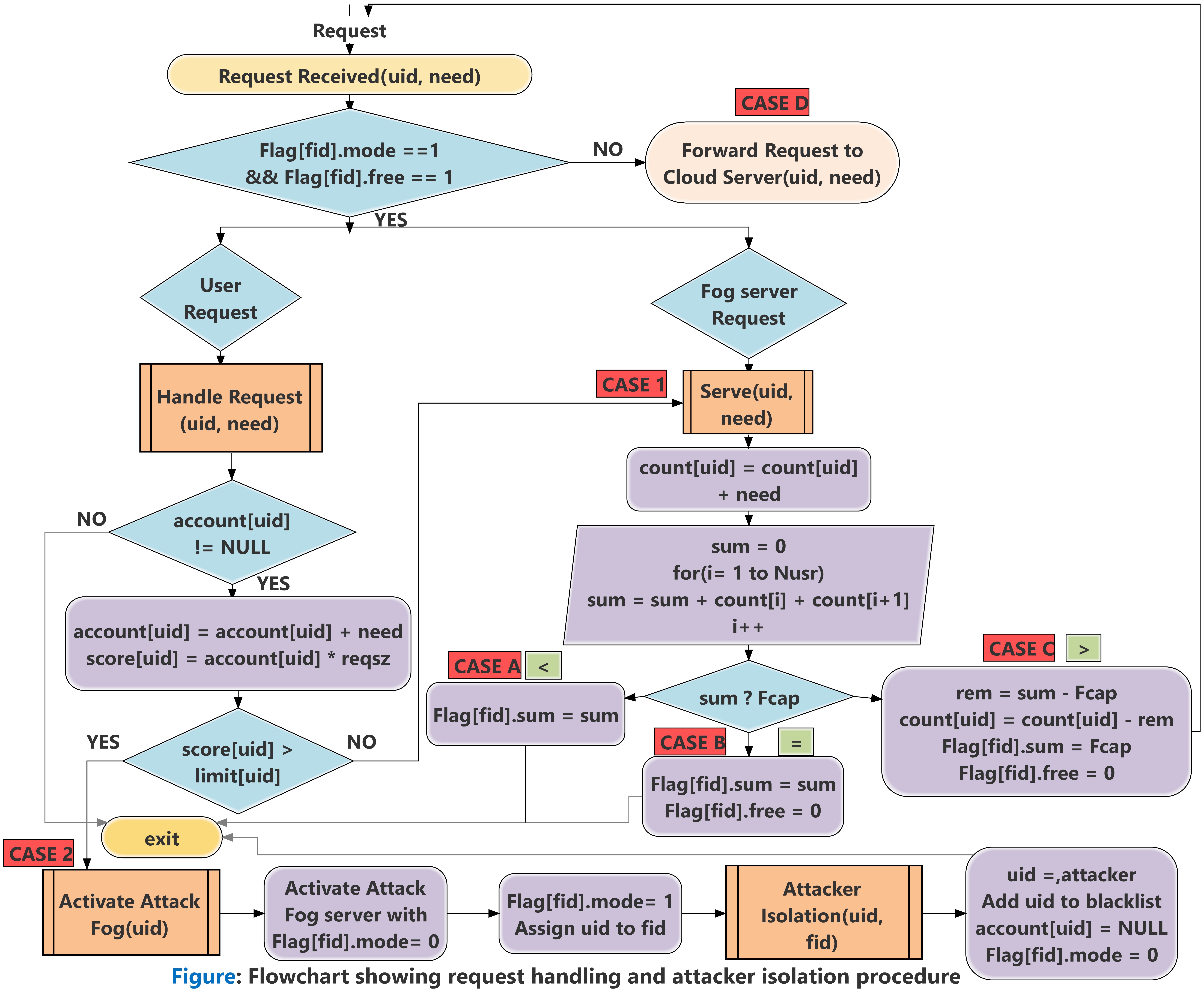} 
\caption{Flowchart showing Request Handling and Attacker Isolation Procedure}
\label{f}
\end{figure}

\par Fig. \ref{f} provides a brief overview with the help of flowchart representing the execution of all cases and the actions taken in all the cases as mentioned in the algorithms.

\section{Result and Analysis}
\par In this paper a conceptual approach \textit{CFPM} is proposed, in which DDoS attack is mitigated at fog layer using migration MTD technique with the aim of minimizing disruption to serving fog servers.
\par Various cases are possible based on the need of the users and current capability of the fog servers which are handled using the proposed algorithms. MATLAB is used to experimentally evaluate the \textit{CFPM} approach.

\subsection{Possible Scenarios}
\par Considering a system with 'n', as the total number of clients where each client is allowed to send maximum 100 requests per second, assuming Reqsize to be 10 Mb. So \emph{max[i]} entry and \emph{limit[i]} for all users "i" would be 100 and 1000 respectively. 
\par  The network parameters that are used during the simulation are listed in Table \ref{tab:tablef}.
\begin{table}[H]
\centering
  \caption{Simulation Parameters }
  \label{tab:tablef}\centering
  \begin{tabularx}{8cm}{|Sl|X|}
    \hline
    \textbf{Parameter} & \textbf{Value} \\
    \hline
    \hline
 thresh & 5 Gbps \\
 \hline
 $Req_{size}$ & 10 Mb \\
 \hline
 $F_{cap}$ & 500 requests/second.\\
 \hline
 max[i] & 100 requests/second  \\
 \hline
 limit[i] & 1000 Mbps  \\
 \hline
 account[i] & Number of requests generated in a second \\
 \hline
 score[i] & calculated \\
 \hline
\end{tabularx}
\end{table}

\par  F1, F2 and F3 are the fog servers present at fog layer in which F1 and F3 are active fog servers while F2 is a sleeping fog server. All the fog servers are having data processing speed as 5 Gbps. So using (2) we can calculate maximum number of requests that a fog server can handle efficiently i.e., $F_{cap}$ = 500.

\par For a user "i" the value of \emph{account[i]} is updated at the time of request generation and further using (1) \emph{score[i]} is calculated. Then \emph{score[i]} is compared to \emph{limit[i]}, which results in following two possible cases 1 and 2: 

\begin{enumerate}

\item[(1)] Case 1: 
\par It is executed when the \emph{score[i]} value is either equal to or less than \emph{limit[i]} value. In this case SERVE() function is called which further compares sum value and $F_{Cap}$ value resulting in four possible cases - A, B, C and D:

\item[(a)] Case A:
\par Fig. \ref{v61} shows the case when the total number of requests served by a fog server is less than capacity of fog server. Fig. \ref{v62} shows the comparison of max[] and account[] values. Table \ref{tab:tableA} shows the final updated values in \emph{Flag[ ]} variable for all respective fog servers.  

\begin{figure}[H]
\centering
\includegraphics[width=2.5in,height=2in]{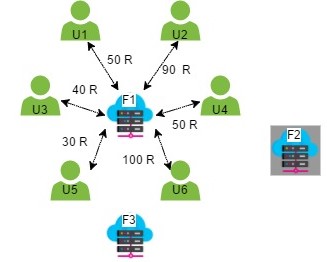} 
\caption{Case A : $Sum < F_{Cap}$}
\label{v61}
\end{figure}

\begin{figure}[H]
\centering
\fbox{\includegraphics[width=2.5in,height=1.2in]{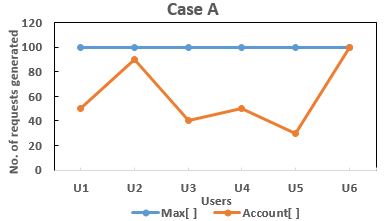} }
\caption{Comparison of Max[ ] and Account[ ]}
\label{v62}
\end{figure}

\begin{table}[htbp]
  \centering
  \caption{Flag variable for case A}
    \begin{tabular}{|l|r|r|r|r|r|r|r|}
    \toprule
    \textbf{Fields} & \multicolumn{7}{c|}{\textbf{Values}} \\
    \midrule
    \midrule
    \textbf{F1.mode} & \multicolumn{7}{c|}{\textbf{1}} \\
    \midrule
    \textbf{F1.free} & \multicolumn{7}{c|}{\textbf{1}} \\
    \midrule
    \textbf{F1.sum} & \textcolor[rgb]{ 1,  0,  0}{\textbf{0}} & \textcolor[rgb]{ 1,  0,  0}{\textbf{50}} & \textcolor[rgb]{ 1,  0,  0}{\textbf{140}} & \textcolor[rgb]{ 1,  0,  0}{\textbf{180}} & \textcolor[rgb]{ 1,  0,  0}{\textbf{230}} & \textcolor[rgb]{ 1,  0,  0}{\textbf{260}} & \textcolor[rgb]{ .329,  .51,  .208}{\textbf{360}} \\
    \midrule
    \midrule
    \textbf{F2.mode} & \multicolumn{7}{c|}{\textbf{0}} \\
    \midrule
    \textbf{F2.free} & \multicolumn{7}{c|}{\textbf{1}} \\
    \midrule
    \textbf{F2.sum} & \textbf{0} &       &       &       &       &       &  \\
    \midrule
    \midrule
    \textbf{F3.mode} & \multicolumn{7}{c|}{\textbf{1}} \\
    \midrule
    \textbf{F3.free} & \multicolumn{7}{c|}{\textbf{1}} \\
    \midrule
    \textbf{F3.sum} & \textbf{0} &       &       &       &       &       &  \\
    \bottomrule
    \bottomrule
    \end{tabular}%
  \label{tab:tableA}%
\end{table}%

\item[(b)] Case B:
\par Fig. \ref{v31} shows the case when the total number of requests served by a fog server is equal to capacity of fog server.

\begin{figure}[h]
\centering
\includegraphics[width=2.5in,height=2in]{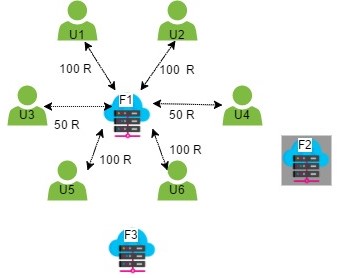} 
\caption{ Case B : $Sum = F_{Cap}$}
\label{v31}
\end{figure}

\par In this case fog server F1 serve users U1,U2,U3,U4,U5 and U6 and updates the value of all the variables. Comparison of max[] and account[] is as shown in Fig. \ref{v32}. After serving it gets fully occupied so it updates its \emph{Flag[F1].free} value to 0 as shown in Table \ref{tab:tableB} depicting that it has reached its \emph{limit} and can no longer serve any user.

\begin{figure}[H]
\centering
\fbox{\includegraphics[width=2.5in,height=1.2in]{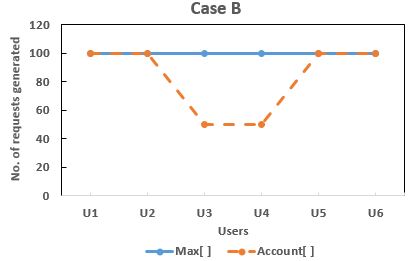} }
\caption{Comparison of Max[ ] and Account[ ]}
\label{v32}
\end{figure}

\begin{table}[htbp]
  \centering
  \caption{Flag variable for case B }
    \begin{tabular}{|l|r|r|r|r|r|r|r|}
    \toprule
    \textbf{Fields} & \multicolumn{7}{c|}{\textbf{Values}} \\
    \midrule
    \midrule
    \textbf{F1.mode} & \multicolumn{7}{c|}{\textbf{1}} \\
    \midrule
    \textbf{F1.free} & \multicolumn{1}{r}{\textcolor[rgb]{ 1,  0,  0}{\textbf{1}}} & \multicolumn{1}{r}{\textcolor[rgb]{ .329,  .51,  .208}{\textbf{0}}} & \multicolumn{1}{r}{} & \multicolumn{1}{r}{} & \multicolumn{1}{r}{} & \multicolumn{1}{r}{} &  \\
    \midrule
    \textbf{F1.sum} & \textcolor[rgb]{ 1,  0,  0}{\textbf{0}} & \textcolor[rgb]{ 1,  0,  0}{\textbf{100}} & \textcolor[rgb]{ 1,  0,  0}{\textbf{200}} & \textcolor[rgb]{ 1,  0,  0}{\textbf{250}} & \textcolor[rgb]{ 1,  0,  0}{\textbf{300}} & \textcolor[rgb]{ 1,  0,  0}{\textbf{400}} & \textcolor[rgb]{ .329,  .51,  .208}{\textbf{500}} \\
    \midrule
    \midrule
    \textbf{F2.mode} & \multicolumn{7}{c|}{\textbf{0}} \\
    \midrule
    \textbf{F2.free} & \multicolumn{7}{c|}{\textbf{1}} \\
    \midrule
    \textbf{F2.sum} & \textbf{0} &       &       &       &       &       &  \\
    \midrule
    \midrule
    \textbf{F3.mode} & \multicolumn{7}{c|}{\textbf{1}} \\
    \midrule
    \textbf{F3.free} & \multicolumn{7}{c|}{\textbf{1}} \\
    \midrule
    \textbf{F3.sum} & \textbf{0} &       &       &       &       &       &  \\
    \bottomrule
    \bottomrule
    \end{tabular}%
  \label{tab:tableB}%
\end{table}%

\item[(c)] Case C:
\par Fig. \ref{v41} shows the case when the total number of requests served by a fog server is greater than the capacity of fog server so the requests are served by some active fog servers collectively. 

\begin{figure}[H]
\centering
\includegraphics[width=3in,height=2.5in]{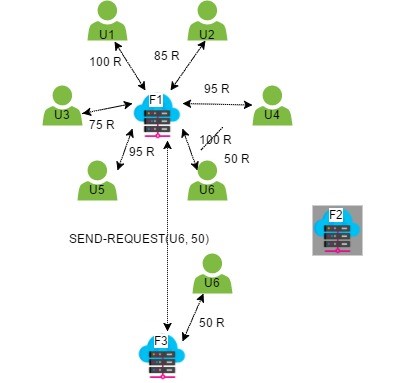} 
\caption{ Case C : $Sum > F_{Cap}$ and request served by fog servers collectively}
\label{v41}
\end{figure}

\par In this case after serving users U1,U2,U3,U4 and U5, fog server F1 serves 450 requests in total and updates the variables accordingly and their comparison is as shown in Fig. \ref{v42}. When U6 generates 100 requests then sum value is incremented to 550 which is greater than the capacity of fog server. So F1 server calculates the remaining need of user U6 which it was unable to serve and update its count[U6] value to the number of requests it has served for user U6.
It then updates its \emph{Flag[F1].sum} value to $F_{Cap}$ and \emph{Flag[F1].free} value to 0 as shown in Table \ref{tab:tableC}, which signifies that it has reached to its \emph{limit}. Further, to serve the remaining need of the user U6 it calls SEND-REQUEST(U6, 50) function and sends request to another available fog server F3. F3 then serve user U6. In this case F1 and F3 fog servers collectively serves the need of user U6 as shown in Fig. \ref{v41}.

\begin{figure}[H]
\centering
\fbox{\includegraphics[width=2.5in,height=1.2in]{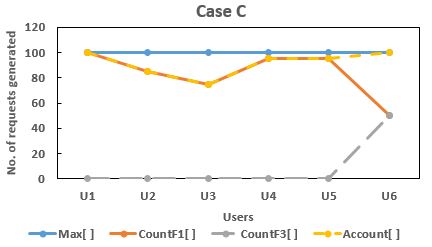} }
\caption{Comparison of various variables for case C}
\label{v42}
\end{figure}

\begin{table}[htbp]
  \centering
  \caption{Flag variable for case C }
    \begin{tabular}{|l|r|r|r|r|r|r|r|}
    \toprule
    \textbf{Fields} & \multicolumn{7}{c|}{\textbf{Values}} \\
    \midrule
    \midrule
    \textbf{F1.mode} & \multicolumn{7}{c|}{\textbf{1}} \\
    \midrule
    \textbf{F1.free} & \multicolumn{1}{r}{\textcolor[rgb]{ 1,  0,  0}{\textbf{1}}} & \multicolumn{1}{r}{\textcolor[rgb]{ .329,  .51,  .208}{\textbf{0}}} & \multicolumn{1}{r}{} & \multicolumn{1}{r}{} & \multicolumn{1}{r}{} & \multicolumn{1}{r}{} &  \\
    \midrule
    \textbf{F1.sum} & \textcolor[rgb]{ 1,  0,  0}{\textbf{0}} & \textcolor[rgb]{ 1,  0,  0}{\textbf{100}} & \textcolor[rgb]{ 1,  0,  0}{\textbf{185}} & \textcolor[rgb]{ 1,  0,  0}{\textbf{260}} & \textcolor[rgb]{ 1,  0,  0}{\textbf{355}} & \textcolor[rgb]{ 1,  0,  0}{\textbf{450}} & \textcolor[rgb]{ .329,  .51,  .208}{\textbf{500}} \\
    \midrule
    \midrule
    \textbf{F2.mode} & \multicolumn{7}{c|}{\textbf{0}} \\
    \midrule
    \textbf{F2.free} & \multicolumn{7}{c|}{\textbf{1}} \\
    \midrule
    \textbf{F2.sum} & \textbf{0} &       &       &       &       &       &  \\
    \midrule
    \midrule
    \textbf{F3.mode} & \multicolumn{7}{c|}{\textbf{1}} \\
    \midrule
    \textbf{F3.free} & \multicolumn{7}{c|}{\textbf{1}} \\
    \midrule
    \textbf{F3.sum} & \textcolor[rgb]{ 1,  0,  0}{\textbf{0}} & \textcolor[rgb]{ .329,  .51,  .208}{\textbf{50}} &       &       &       &       &  \\
    \bottomrule
    \bottomrule
    \end{tabular}%
  \label{tab:tableC}%
\end{table}%

\item[(d)] Case D: 
\par Fig. \ref{v51} shows the case when the total number of requests served by a fog server is greater than the capacity of fog server and all the active fog servers collectively are unable to serve the requests of the user completely. So to handle the case request is forwarded to cloud server. Fig. \ref{v52} shows the comparison of all the variables for all users. Final value of \emph{Flag[ ]} variable for all fog servers is shown in Table \ref{tab:tableD} .

\begin{figure}[H]
\centering
\includegraphics[width=4in,height=3.8in]{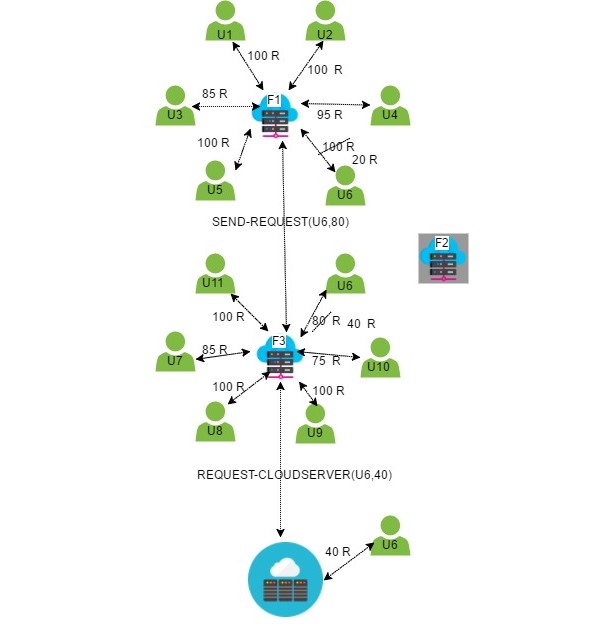} 
\caption{Case D : $Sum > F_{Cap}$ and request served by cloud server}
\label{v51}
\end{figure}

\begin{figure}[H]
\centering
\fbox{\includegraphics[width=2.5in,height=1.2in]{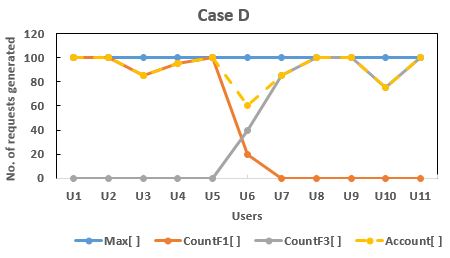} }
\caption{Comparison of various variables for case D}
\label{v52}
\end{figure}

\begin{table}[htbp]
  \centering
  \caption{Flag variable for case D}
    \begin{tabular}{|l|r|r|r|r|r|r|r|}
    \toprule
    \textbf{Fields} & \multicolumn{7}{c|}{\textbf{Values}} \\
    \midrule
    \midrule
    \textbf{F1.mode} & \multicolumn{7}{c|}{\textbf{1}} \\
    \midrule
    \textbf{F1.free} & \multicolumn{1}{r}{\textcolor[rgb]{ 1,  0,  0}{\textbf{1}}} & \multicolumn{1}{r}{\textcolor[rgb]{ .329,  .51,  .208}{\textbf{0}}} & \multicolumn{1}{r}{} & \multicolumn{1}{r}{} & \multicolumn{1}{r}{} & \multicolumn{1}{r}{} &  \\
    \midrule
    \textbf{F1.sum} & \textcolor[rgb]{ 1,  0,  0}{\textbf{0}} & \textcolor[rgb]{ 1,  0,  0}{\textbf{100}} & \textcolor[rgb]{ 1,  0,  0}{\textbf{200}} & \textcolor[rgb]{ 1,  0,  0}{\textbf{285}} & \textcolor[rgb]{ 1,  0,  0}{\textbf{380}} & \textcolor[rgb]{ 1,  0,  0}{\textbf{480}} & \textcolor[rgb]{ .329,  .51,  .208}{\textbf{500}} \\
    \midrule
    \midrule
    \textbf{F2.mode} & \multicolumn{7}{c|}{\textbf{0}} \\
    \midrule
    \textbf{F2.free} & \multicolumn{7}{c|}{\textbf{1}} \\
    \midrule
    \textbf{F2.sum} & \textbf{0} &       &       &       &       &       &  \\
    \midrule
    \midrule
    \textbf{F3.mode} & \multicolumn{7}{c|}{\textbf{1}} \\
    \midrule
    \textbf{F3.free} & \multicolumn{1}{r}{\textcolor[rgb]{ 1,  0,  0}{\textbf{1}}} & \multicolumn{1}{r}{\textcolor[rgb]{ .329,  .51,  .208}{\textbf{0}}} & \multicolumn{1}{r}{} & \multicolumn{1}{r}{} & \multicolumn{1}{r}{} & \multicolumn{1}{r}{} &  \\
    \midrule
    \textbf{F3.sum} & \textcolor[rgb]{ 1,  0,  0}{\textbf{0}} & \textcolor[rgb]{ 1,  0,  0}{\textbf{85}} & \textcolor[rgb]{ 1,  0,  0}{\textbf{185}} & \textcolor[rgb]{ 1,  0,  0}{\textbf{285}} & \textcolor[rgb]{ 1,  0,  0}{\textbf{360}} & \textcolor[rgb]{ 1,  0,  0}{\textbf{460}} & \textcolor[rgb]{ .329,  .51,  .208}{\textbf{500}} \\
    \bottomrule
    \bottomrule
    \end{tabular}%
  \label{tab:tableD}%
\end{table}%

\par 
\item[(2)] Case 2:
\par This case is executed when the \emph{score[i]} value is greater than \emph{limit[i]} value for any user "i". Fig. \ref{v71} shows how the attacker isolation procedure works. Fig. \ref{v72} shows the comparison of all the variables for the users and updated blacklist.
\par In this U2 is generating 200 requests that results in its \emph{score[U2]} value as 2000 which is greater than its \emph{limit[U2]} value. So ACTIVATE-ATTACKFOG function is called, which activates a sleeping fog server F2 also named as attack fog server by changing \emph{Flag[F2].mode} bit value to 1. F2 fog server declares U2 as attacker and add U2 to blacklist. It then updates \emph{account[U2]} value to NULL, which act as a notification for all the fog servers that no one has to serve user U2 as it is observed as an attacker.
\par After this attacker U2 is assigned to the attack fog server and again it is send to the sleeping mode by changing its \emph{Flag[F2].mode} bit value to 0. This results in the isolation of attacker U2 from the system. Same procedure is executed for user U5 also. Updations made to the \emph{Flag[ ]} variable are shown in Table \ref{tab:table2}.

\begin{figure}[h]
\centering
\includegraphics[width=3in,height=2.5in]{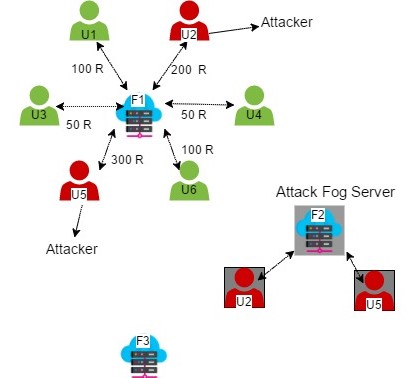} 
\caption{Case 2 : Attacker Isolation}
\label{v71}
\end{figure}

\begin{figure}[h]
\centering
\fbox{\includegraphics[width=2.5in,height=1.2in]{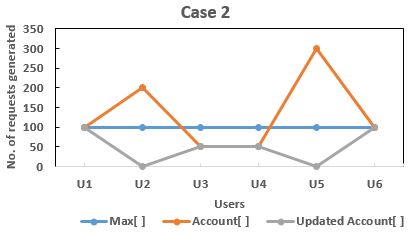} }
\caption{Comparison of the variables for case 2}
\label{v72}
\end{figure}

\begin{table}[htbp]
  \centering
  \caption{Flag variable for case 2}
    \begin{tabular}{|l|r|r|r|r|r|r|r|}
    \toprule
    \textbf{Fields} & \multicolumn{7}{c|}{\textbf{Values}} \\
    \midrule
    \midrule
    \textbf{F1.mode} & \multicolumn{7}{c|}{\textbf{1}} \\
    \midrule
    \textbf{F1.free} & \multicolumn{7}{c|}{\textbf{1}} \\
    \midrule
    \textbf{F1.sum} & \textcolor[rgb]{ 1,  0,  0}{\textbf{0}} & \textcolor[rgb]{ 1,  0,  0}{\textbf{100}} & \textcolor[rgb]{ 1,  0,  0}{\textbf{150}} & \textcolor[rgb]{ 1,  0,  0}{\textbf{200}} & \textcolor[rgb]{ .329,  .51,  .208}{\textbf{300}} & \textcolor[rgb]{ 1,  0,  0}{} & \textcolor[rgb]{ .329,  .51,  .208}{} \\
    \midrule
    \midrule
    \textbf{F2.mode} & \multicolumn{1}{r}{\textbf{0}} & \multicolumn{1}{r}{\textbf{1}} & \multicolumn{1}{r}{\textbf{0}} & \multicolumn{1}{r}{} & \multicolumn{1}{r}{} & \multicolumn{1}{r}{} &  \\
    \midrule
    \textbf{F2.free} & \multicolumn{7}{c|}{\textbf{1}} \\
    \midrule
    \textbf{F2.sum} & \textbf{0} &       &       &       &       &       &  \\
    \midrule
    \midrule
    \textbf{F3.mode} & \multicolumn{7}{c|}{\textbf{1}} \\
    \midrule
    \textbf{F3.free} & \multicolumn{7}{c|}{\textbf{1}} \\
    \midrule
    \textbf{F3.sum} & \textbf{0} & \textcolor[rgb]{ 1,  0,  0}{} & \textcolor[rgb]{ 1,  0,  0}{} & \textcolor[rgb]{ 1,  0,  0}{} & \textcolor[rgb]{ 1,  0,  0}{} & \textcolor[rgb]{ 1,  0,  0}{} & \textcolor[rgb]{ .329,  .51,  .208}{} \\
    \bottomrule
    \bottomrule
    \end{tabular}%
  \label{tab:table2}%
\end{table}%

\end{enumerate}

\begin{figure}[h]
\centering
\fbox{\includegraphics[width=3in,height=1.2in]{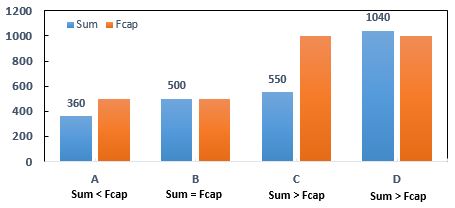} }
\caption{Comparison of Cases A,B,C and D}
\label{v20}
\end{figure}

Fig.\ref{v20} shows the comparison of Cases A,B,C and D under Case 1 with respect to Sum and $F_{Cap}$ and number of fog servers needed (for case C, two fog servers would be needed and for case D even two fog servers would not suffice. Therefore remaining need is forwarded to cloud server).

\subsection{Evaluation}
\par
\begin{figure}[h]
\centering
\fbox{\includegraphics[width=3in,height=1.7in]{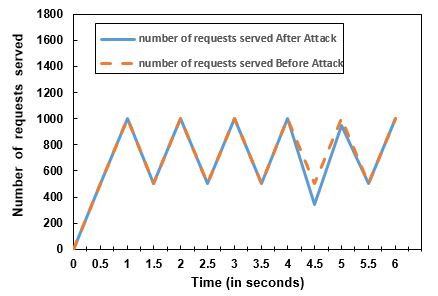} }
\caption{Evaluation of the system before and after attack }
\label{v8}
\end{figure}

 Fig. \ref{v8} shows the evaluation of the system on implementation of \textit{CFPM} approach before and after attack. A fog server can handle 500 requests per second so in one second 1000 requests can be served by two fog servers. In case there is no attacker present, after every second same number of requests are handled.

\par  But if there is an attacker present in the system, say the attacker arrives at 4th second, then the fog server detect it by comparing certain values and call ACTIVATE-ATTACKFOG function. 
\par Detection and function calling will take very less time (fraction of seconds), so for only that amount of time there would be a minor decrease in the number of requests served by fog server. After that fog server can process the request in the usual manner. As after detection of attacker, attack fog server is responsible for handling the attacker without any involvement of the detecting fog server. This evaluation proves that the proposed approach \textit{CFPM} is very effective.

\section{Conclusion}
\par For mitigating DDoS attacks in the cloud, different approaches were proposed in the literature. Fog computing and MTD has been used separately to address the security issues in cloud.   MTD has never been applied to fog paradigm. After evaluating the effectiveness of MTD it has been observed that MTD can be applied to decentralized cloud infrastructure such as fog. This paper presents a proactive novel approach \textit{CFPM}, that mitigates DDoS attacks in cloud using fog layer with applied migration MTD technique. \textit{CFPM} detects and isolate the attacker based on certain network parameters at fog layer. By using migration MTD technique it migrates the attacker to attack fog server minimizing disruption to serving fog servers and requiring no involvement of cloud server. Attack fog server is an activated fog server which after migration of attacker again goes into the sleep mode resulting in the isolation of the attacker. \textit{CFPM} approach is analyzed under different cases  and performance of the system before and after attack is evaluated. \textit{CFPM} uses the idea of load balancing to prevent fog servers from getting overloaded.

\bibliography{Rf}
\bibliographystyle{IEEEtran}

\end{document}